\documentclass[prl,aps,showpacs,superscriptaddress,twocolumn]{revtex4-1}
\usepackage{amsmath,amssymb,amsfonts,latexsym,fancyhdr,graphicx,epstopdf,times,txfonts}

\usepackage{graphicx}
\usepackage{verbatim}
\makeatletter

\usepackage{color}

\begin{document}

\title{Thermodynamic meaning and power of non-Markovianity}

\author{Bogna Bylicka$^{1,2}$, Mikko Tukiainen $^3$, Dariusz Chru\'sci\'nski$^2$, Jyrki Piilo$^3$, and Sabrina Maniscalco$^{3}$\\
$^1$ICFO-Institut de Ciencies Fotoniques, Mediterranean Technology Park, 08860 Castelldefels (Barcelona), Spain\\
$^2$Institute of Physics, Faculty of Physics, Astronomy and Informatics, Nicolaus Copernicus University, Grudzi\c{a}dzka 5/7, 87--100 Toru\'n, Poland\\
$^3$Turku Centre for Quantum Physics, Department of Physics and Astronomy,
University of Turku, FI-20014, Turun Yliopisto, Finland}

%
%
%

\date{\today}

\begin{abstract}
We establish a connection between non-Markovian memory effects and thermodynamical quantities such as work. We show how memory effects can be interpreted as revivals of work that can be extracted from a quantum system. We prove that non-Markovianity may allow an increase in the extractable work even when the entropy of the system is increasing. Our results have important implications both in quantum thermodynamics and in quantum information theory. In the former context they pave the way to the understanding of concepts like work in a non-Markovian open system scenario. In the latter context they lead to interesting consequences for quantum state merging protocols in presence of noise.

\end{abstract}


\maketitle

\section{Introduction}

The connection between thermodynamics and information theory, expressed by Landauer's principle, is a milestone of the physics of the last century. According to this principle, the erasure of information stored in a system requires an amount of work proportional to the entropy of the system.
The natural framework to discuss thermodynamics at the quantum level is the theory of open quantum systems. A number of recent results have shown that memory effects arising from strong system-environment correlations may lead to information back-flow, hence prolonging the life of quantum properties. Open systems exhibiting such behaviour are known as non-Markovian. 
The relation between non-Markovianity and quantum thermodynamics has been until now largely unexplored. 
Here we establish this missing link by means of Landauer's principle. We show that memory effects control the amount of work that one can extract from an open quantum system. Hence, the work extraction can be optimised via reservoir engineering techniques.

Landauer's erasure principle states that, in order to perform irreversible operations on a system, such as erasing a bit of information, a certain amount of work must be performed. This is known as the work of erasure \cite{Landauer}. The very same  principle holds in the quantum regime carrying far-reaching consequences \cite{Koji,Lidia}. As noticed in Ref. \cite{Lidia}, indeed, in a quantum setting the presence of a quantum memory correlated to the system leads to the possibility of extracting work while erasing information on the system. This possibility crucially relies on the persistence of quantum correlations between the system and the quantum memory. One would therefore expect that, due to the usually deleterious effect of the environment acting on either the system or the memory, this uniquely quantum feature should only survive for a very short time. In other words we expect that, after the initial preparation of a quantum correlated state between system and memory, the ability to extract work by erasure will characterise only a short initial transient.

One of the main results of this Article is to prove that this needs not be true for non-Markovian systems \cite{NMQJ,NMBLP,NMCV,NMWolf,NMRHP,NMLuo,NMFrancesco,NMNatPhys,NMSabDarek}. 
We will show how memory effects are directly connected to the temporal behaviour of the work of erasure and are therefore instrumental to the persistence of work extraction. This is important also in the ongoing discussion on the usefulness of non-Markovianity for quantum technologies \cite{NMQCAP,NMRQKD,NMRMetrology,NMRTeleportation}, since our results prove for the first time that memory effects are useful for work extraction purposes. 

\section{Landauer principle in the open system scenario.} 
The extension of Landauer's principle to quantum systems, formulated by Lubkin, states that the work performed on a quantum system $S$ to erase classical information stored in it, is connected to the system's entropy via the relation
\begin{equation}
W(S)=H(S) \, kT  \ln 2, \label{eq:Wclas}
\end{equation}
where $H(S)$ is the von Neumann entropy quantifying our lack of knowledge on the system, $k$ is Boltzmann constant, and $T$ is the temperature of the reservoir used for the optimal erasure  process.

When one tries to ascertain properties of the system, however, one is naturally led to introduce an observer $O$ which performs measurements on $S$. In this framework the \lq\lq amount of information\rq\rq  stored in a system is observer-dependent. One should therefore replace the von Neumann entropy $H(S)$ in Eq. ({\ref{eq:Wclas}}) with the conditional entropy $H(S|O)$.
The fully quantum scenario is described by a situation in which the observer possesses a quantum memory $Q$. In Ref. \cite{Lidia} it is shown that, in the thermodynamic limit, the work of erasure is given by
\begin{equation}
W(S|Q) = H(S|Q) \, kT  \ln2, \label{eq:WSQ}
\end{equation}
and the extractable work, that is the work that one can extract from an $n$-qubit  system, is
\begin{equation}
W_{ex} = [n-H(S|Q)] \, kT \ln2, \label{eq:WEXT}
\end{equation}
where $H(S|Q)=H(SQ)-H(Q)$ is the conditional entropy, i.e. the entropy of the system conditioned on the memory, with $H(SQ)$  and $H(Q)$ the von Neumann entropies of the combined system-memory and of the memory, respectively. If the system and the memory are prepared in a quantum correlated state, such as maximally entangled state, the conditional entropy may be negative. When this happens one can extract work while erasing information on the system. This is a genuinely  quantum phenomenon with no classical counterpart.

We will now extend this analysis to the case in which either the system or the memory is subjected to the action of the environment. This will give us a basic understanding of the robustness of the quantum-enhanced work extraction to realistic and unavoidable sources of noise. The two situations are pictorially illustrated in Fig. 1. 
\begin{figure}
\includegraphics[width=8cm]{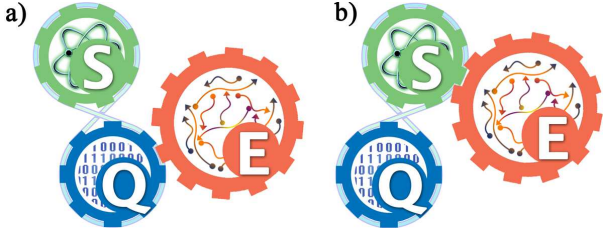} 
\caption{(Colors online) Illustration of the two physical scenarios considered in the Article: a) the open quantum memory and b) the open quantum system.}
\end{figure}
Both the work of erasure and the extractable work are in these scenarios time-dependent. One would expect that, due to the presence of the environment, the work of erasure would increase in time while the extractable work would decrease. Let us indicate with 
\begin{eqnarray}
\Delta W_{ex}(t_1,t_2) &=& W_{ex}(t_2) - W_{ex}(t_1) \nonumber \\
&=& [H(S|Q)_{t_1}-H(S|Q)_{t_2}] \, kT \ln2, \label{eq:DWext}
\end{eqnarray}
with $t_2 \ge t_1$, the variation in time of the extractable work, and with $\Delta W(t_1,t_2)= - \Delta W_{ex}(t_1,t_2)$ the change in the work of erasure.

It is worth stressing that the processes of work extraction or erasure \cite{Aberg,HorodeckiOp,Popescu1} are not directly considered in our scheme and are assumed to take place at a given time $t$, after the system has interacted with the environment. The optimal process to extract the amount of work given by Eqs. (\ref{eq:WSQ})-(\ref{eq:WEXT}) is described in \cite{Lidia}. Our main goal is to understand how the environment affects, as time evolves, the ability to extract work from the system. By using reservoir engineering techniques \cite{VWC,multiparticleendyn,DDengineer} one may then be able to minimize the detrimental effects of the environment. As we will see, memory effects will be the key ingredient to accomplish this goal. We begin by considering the case of an open quantum memory.

{\it Open quantum memory:} The action of the environment on the quantum memory is described in terms of a dynamical map $\Phi_t$, that is a $t$-parametrized family of completely positive and trace preserving (CPTP) maps. 
We indicate the conditional entropy in Eq. (\ref{eq:DWext}) with $H(S|Q_{t})$, with $Q_t=\Phi_t(Q)$, to emphasize that only $Q$ is evolving due to the coupling to the environment. This quantity is just the negative of the coherent information, i.e., $I(S\rangle Q_t) =-H(S|Q_t)$, measuring the correlations between the system and the memory \cite{Wilde}. Hence,
\begin{equation}
\Delta W_{ex}(t_1,t_2) =[I(S\rangle Q_{t_2})-I(S\rangle Q_{t_1})] \, kT \ln2. \label{eq:Idontknow}
\end{equation}
For divisible dynamical maps, namely when $\Phi_{t} = \Phi_{t,s} \Phi_{s}$, with $s\le t$ and $\Phi_{t,s} $ CPTP, the data processing inequality implies, that 
\begin{equation}\label{QDP}
I(S\rangle Q_{t_2}) \leq I(S\rangle Q_{t_1}), \quad \mbox{for} \ t_1 \leq t_2.
\end{equation}
In this case, the extractable work monotonically decreases in time and the work of erasure monotonically increases, as one would intuitively expect. 

{\it Open quantum system:} Let us now investigate what happens when the environment acts on the system rather than on the memory. The extractable work is now given by
\begin{eqnarray}
W_{ex}(t)&=&[n-H(S_{t}|Q)]\, kT \ln2 \nonumber \\
& =& [n- H(S_t)+I(S_t : Q)]\, kT \ln2, 
\end{eqnarray}
where $H(S_{t}|Q)$ indicates the conditional entropy when the system is coupled to the environment, i.e., $S_t=\Phi_t(S)$, and ${I(S_t: Q) = H(S_t) + H(Q) - H(S_t Q)}$ is the mutual information at time $t$, quantifying the amount of information that $S_t$ and $Q$ share \cite{Wilde}.

The change in the extractable work can be recast in the form
\begin{equation}
\Delta W_{ex}(t_1,t_2)=[-\Delta H(S_t)+ \Delta I(S_t:Q)] \, kT \ln2, \label{eq:openSWex}
\end{equation}
where $\Delta H(S_t)=H(S_{t_2})-H(S_{t_1})$ and $\Delta I(S_t:Q)=  I(S_{t_2}:Q) -  I(S_{t_1}:Q)$, with $t_2 \ge t_1$.
Once again, for divisible dynamics, 
\begin{equation}
I(S_{t_2}:Q) \leq I(S_{t_1}:Q), \quad \text{for} \ t_1\leq t_2,
\end{equation}
which implies that an increase in extractable work requires a decrease in entropy of the system. This meets our intuition, since it amounts to saying that the more we know about the system, namely the closer it is to a pure state, the more work we can extract from it. Equivalently, the closer the system is to a maximally mixed state, the smaller is the extractable work. As we will see in the following, however, non-Markovian memory effects may defy this intuition.


We note in passing that for unital dynamical maps the change in the entropy of the system vanishes when the initial system-memory (2 qubit system) is prepared in a maximally entangled state. In this case we obtain exactly the same result of the open quantum memory scenario, that is, the variation in the extractable work is proportional to the variation in the coherent information, see Eq. (\ref{eq:Idontknow}), since $\Delta I(S_t:Q) = \Delta I(S_t\rangle Q)$.

\section{Thermodynamic meaning of non-Markovianity}
The most general form of open system evolution is described by dynamical maps which may violate the divisibility property while still being CPTP. When this happens, the coherent information may temporarily increase for certain time intervals. This, in turn, generally leads to a partial recovery of extractable work or, equivalently, to a decrease in the work of erasure. 

Recent research on open quantum systems has highlighted the fact that memory effects can be associated to different manifestations of non-Markovianity which, in turn, are revealed by different physical quantities. Several non-Markovianity measures capturing this plethora of phenomena have been introduced and used to describe the information back-flow and the recoherence \cite{NMQJ,NMBLP,NMCV,NMWolf,NMRHP,NMLuo,NMFrancesco,NMNatPhys,NMSabDarek}. A connection between memory effects and thermodynamical quantities, however, was until now missing. Equations (\ref{eq:Idontknow}) and (\ref{eq:openSWex}) bridge this gap by showing how memory effects are indeed directly related to work.

\begin{figure}
\includegraphics[width=8cm]{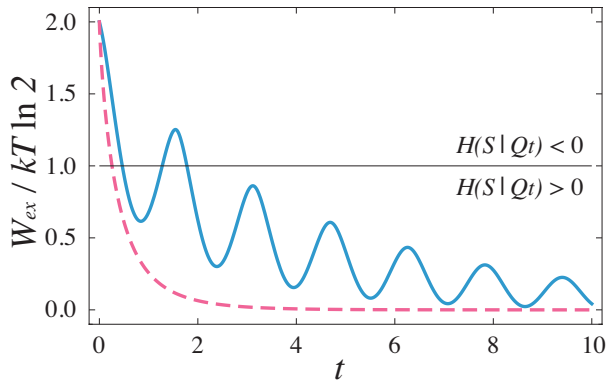} 
\caption {(Colors online) Time evolution of the extractable work $W_{ex}/ kT \ln 2$ for one qubit system. The blue solid line corresponds to a Pauli channel model with decay rates $\gamma_1(t)=\gamma_2(t)= \lambda/2$, and $\gamma_3(t)= \omega \tan(\omega t)/2$, with $\lambda =0.1$ and $\omega=2$ (a.u.). The red dashed line corresponds to a Pauli channel model with decay rates $\gamma_1(t)=\gamma_2(t)= \lambda/2$, and $\gamma_3(t)= - \omega \tanh(\omega t)/2$, with $\lambda =1$ and $\omega=0.5$ (a.u.).}
\end{figure}

Until now our results are model-independent, hence fully general, but formal. The key relevant question is: are there physical models of realistic environments in which the thermodynamic power of non-Markovianity can be observed and tested? We will positively answer to this question by considering three exemplary cases, namely a qubit coupled to a classical environment (Pauli channel), a spin environment (Ising model) and a bosonic environment (amplitude damping channel). 

In our first example we study the time evolution of the extractable work for time-dependent Pauli channels. Pauli channels describe depolarising noise acting independently along the $x$, $y$, and $z$ directions with time-dependent rates $\gamma_1(t)$, $\gamma_2(t)$ and $\gamma_3(t)$, respectively. In this case, due to the unitality of Pauli channels, the behaviour of extractable work in time is described by Eq. (\ref{eq:Idontknow}) both in the open quantum memory and in the open quantum system scenario. In Fig. 2 we compare two different open dynamics. The corresponding behaviours of the extractable work are given by the solid blue line and the dashed red line in Fig. 2. In both cases the dynamical map is non-divisible, but only in one case oscillations of extractable work do take place. When these oscillations occur, the uniquely quantum feature of negative conditional entropy may show revivals even after becoming positive due to noise induced by the environment. For detailed calculations of the time evolution of the extractable work we refer to the Supplementary Material.

This example illustrates that, when defined in terms of coherent (or mutual) information \cite{NMLuo}, the physical interpretation of non-Markovianity as information back-flow carries along a powerful connection to revivals of the extractable work. We can therefore interpret memory effects in terms of a quantity with a clear physical meaning, that is work, rather than the somewhat elusive concept of information. Equation (\ref{eq:Idontknow}), indeed, straightforwardly illustrates that memory effects are manifested as time revivals of work.

\section{ Non-Markovian enhancement of work extraction by erasure} 
As the second example, we consider a physical system consisting of a central spin coupled to a spin chain. More precisely we focus on the Ising model in a transverse field \cite{Zanardi}. Details on the Hamiltonian and the exact solution for the spin system are given in the Supplementary Material. This systems exhibits a quantum phase transition between ferromagnetic state and paramagnetic state when the parameter $\lambda^*$, measuring the ratio between the strength of the transverse field and the coupling between the environmental spins equals unity. In Ref. \cite{PinjaJohn} it has been shown that the central spin dynamics is non-Markovian for any value of $\lambda^*$, except at phase transition. This is therefore an ideal system to study the modification of the dynamics induced by non-Markovianity as we can control the non-Markovian character via the parameter $\lambda^*$. We notice that, also in this case, the dynamical map is unital.  

Once again, due to unitality, the extractable work by erasure is just proportional to the coherent information both in the open system and the open memory scenarios.
Fig. 3 shows the behaviour of extractable work for three exemplary values of $\lambda^*$. For $\lambda^* = 1$ (black solid line) the extractable work quickly decreases to unity signalling that the erasure work goes to zero after a short transient time. In this case the spin environment is at quantum phase transition and the dynamics of the central spin is Markovian. 

For $\lambda^* = 0.1$ (green short-dashed line), after decreasing to unity, the extractable work presents oscillations with maximum amplitude damped in time. These oscillations, corresponding to revivals of the coherent information, characterize the dynamics when the environment is in its ferromagnetic ground state. Finally, for $\lambda^*=1.9$ (orange long-dashed line)  oscillations in the extractable work are also present while this quantity remains larger than unity at all the times, meaning that work extraction from the erasure is always possible. This corresponds to the case, in which the environment is in a paramagnetic state. This example illustrates how non-Markovian memory effects allow to retain or regain extractable work, leading to a thermodynamic enhancement compared to the Markovian case.

\begin{figure}
\includegraphics[width=8cm]{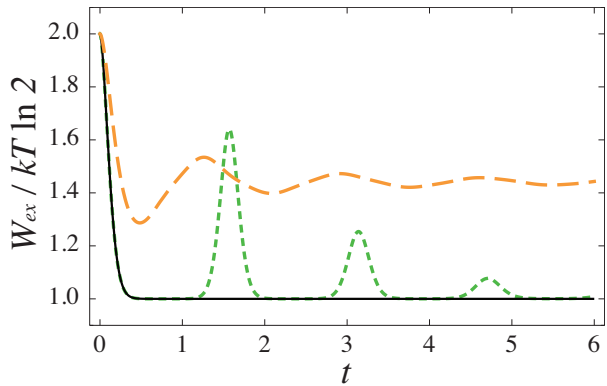} 
\caption{(Colors online) Time evolution of the extractable work of one qubit for an Ising model in a transverse field. We compared the dynamics for $N=4000$ spins and three values of renormalized transverse field $\lambda$: $\lambda^* =1$ (black solid line), $\lambda^* = 0.1$ (green short-dashed line) and $\lambda^* = 1.9$ (orange long-dashed line). }
\end{figure}

\section{ The power of quantum correlations}
The third example highlights the interesting situation in which memory effects are responsible for a counter-intuitive behaviour of the extractable work. In the open quantum system scenario the change in the extractable work is given by Eq. (\ref{eq:openSWex}), namely it is the difference between the change in mutual information and the change in entropy. When the open system dynamics is Markovian, the mutual information monotonically decreases. Hence, in this case an increase of the extractable work can only occur when the entropy of the system decreases. Non-Markovian dynamics, however, allows for revivals of the extractable work even when the entropy of the system is constant or increasing. Let us discuss these two cases separately.

We begin by considering non-Markovian dephasing along an arbitrary direction, i.e., a Pauli channel with two null dephasing rates and the third one time-dependent. For a generic initial state of the system one can easily prove that the entropy change is either null or decreasing whenever the mutual information increases. Therefore the best scenario for memory-induced revivals of extractable work  corresponds to the case in which the change in entropy is zero. This is the situation occurring, e.g., in the Ising model in a transverse field, see Fig. 3. 

Let us now focus on the more counter-intuitive situation in which the increase in the coherent information is greater than the increase in the system's entropy, that is, we get an overall increase in the work extraction even though our knowledge of the system decreases. Since we assume that the initial system-memory state is maximally entangled, this situation may occur only for non-unital open system dynamics. Fig. 4 illustrates the dynamics of the extractable work, as defined in Eq. (\ref{eq:openSWex}), the mutual information and the system entropy for an amplitude damping photonic band gap model. Once again all quantities can be calculated analytically; see the Supplementary Material.

In the figure, the green shaded regions highlight time intervals in which memory effects are so dominant to induce a partial recovery of the extractable work even if the system entropy increases. This apparent contradiction is in fact resolved when one thinks that, in this setting, what counts is the amount of information on the system possessed by an observer with a quantum memory. Therefore, it is clear that an increase in the system-memory correlations, indicated by the behaviour of the mutual information, plays a key role in determining the extractable work.

\begin{figure}
\includegraphics[width=8cm]{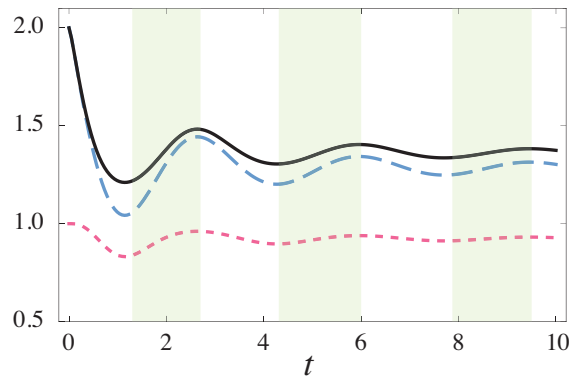} 
\caption{(Colors online) Time evolution of the mutual information (blue long-dashed line), system entropy (red short-dashed line), and extractable work (solid black line), for the amplitude damping photonics band gap model, with detuning from the band gap edge frequency $\delta=-1$ (a.u.).}
\end{figure}

\section{  Application to quantum information theory} 

We conclude by noticing that our results carry along interesting consequences when applied to one of the most important protocols in quantum information theory, namely quantum state merging.
In quantum state merging two parties, Alice and Bob, share a common state $\rho_{AB}$, and Bob has to reproduce this state having unlimited access to local operations and classical communication.
The quantum cost of the protocol, i.e., the minimal amount of Bell pairs that Alice and Bob need to share in order to complete the task, is given by the conditional entropy $H(A|B)$ \cite{StateM}. We see that the crucial quantity here is the same as in the work extraction protocol. Hence, one can think, analogously to what we did in the work extraction protocol, in terms of the degradation of the quantum cost when either Alice or Bob are subjected to an environment. 

The same lines of thought that we used above to illustrate the thermodynamic power of non-Markovianity lead us to conclude that reservoir engineering can be used to induce memory effects hence lowering the quantum cost of the state merging protocol in presence of unavoidable noise. Moreover, it is known that quantum state merging is a primitive, i.e. it can generate all the other building blocks of quantum Shannon theory \cite{StateMMother}. Hence reservoir memory effects are potentially beneficial for any quantum information protocol as they can improve their efficiency.

\section{Conclusions}
The interpretation of non-Markovian memory effects in terms of revivals of either the erasure work or the extractable work unveils a fundamental link between the theory of open quantum systems and quantum thermodynamics. Such a connection also gives a powerful answer to the question of the potential usefulness of non-Markovianity, when we think in terms of reservoir engineering. One of the current challenges of open quantum system theory is the development of a resource theory of non-Markovianity. It is worth underlining that both in quantum thermodynamics and in quantum information theory resource theories have been successfully developed \cite{Jonathan}. Our approach may therefore pave the way to establishing, in the most general framework, if and in which way memory effects can be considered as a resource.

\section*{Acknowledgements}
S.M. and B.B. acknowledge useful discussions with O. Dahlsten and J. Goold. B.B. acknowledges financial support from the Polish Ministry of Science and Higher Education (The "Mobility
Plus'' Program grant no 1107/MOB/2013/0), the Spanish project
FOQUS and the Generalitat de Catalunya (SGR 875). S.M. and J.P. acknowledge financial support from the EU Collaborative project QuProCS (Grant Agreement 641277) and the Magnus Ehrnrooth Foundation.  J.P. also acknowledges the support from Jenny and Antti Wihuri Foundation. M.T. acknowledges financial support from the University of Turku Graduate School (UTUGS). D.C. was partially supported by the National Science Center project
DEC-2011/03/B/ST2/00136.

\section*{Contributions}
B.B. performed the theoretical analysis with the help of M.T. and with comments and input from all authors. S.M., B.B. and M.T. wrote the paper with feedback from D.C. and J.P.
All authors contributed to the design of the research. S.M. initiated and supervised the research project.

\section*{Competing financial interests}
The authors declare no competing financial interests.

\end{document}



\title{Supplementary Material}

\maketitle

In this Supplementary Material we present in detail the mathematical description of the physical models discussed in the main Article.
In all the considered examples we take as an initial state of the system $S$ and quantum memory $Q$ the maximally entangled two qubit state, $|SQ\rangle = \frac{1}{\sqrt{2}}\left(|10\rangle-|01\rangle\right)$. The motivation for such choice comes form the fact that the maximally entangled states are optimal states from the point of view of the work by erasure protocol \cite{Lidia}.

In the Article we investigate two scenarios, when either the qubit of quantum memory or the qubit of system is subjected to environmental noise, hence undergoing a non-unitary evolution. In general, these two situations have different implications on the behaviour in time of the extractable work (see the Eqs. (5) and (8) of the main Article, respectively). However, if the dynamical map is unital and the reduced state of the system is maximally mixed, as it is in fact the case in all studied examples, then the two scenarios are equivalent and provide a direct link between the revivals of the extractable work and the revivals of coherent information of the dynamical map.

Two of the analysed physical models, Pauli channels and Ising model, are unital maps, which allows us to clarify the relation between divisibility, revivals of the coherent information and the work extraction. The last example, photonic band gap amplitude damping, is a non-unital map. Hence, we can study the more complex situation described in the Article's Eq. (8), where also the system entropy is evolving, and witness the surprising behaviour of the extractable work in time.

\subsection{Pauli Channel}

As the first example of a non-Markovian qubit channel, we consider a Pauli channel with dynamical map of the form
\begin{eqnarray}
 \Phi_t(\rho) = \sum_{\alpha=0}^3 p_{\alpha}(t)\, \sigma_{\alpha} \rho \sigma_{\alpha},
\end{eqnarray}
where $0\leq p_{\alpha} \leq 1$ satisfy $\sum_{\alpha} p_{\alpha} =1$, and $\sigma_{\alpha}$ are the Pauli matrices ($\sigma_0 = \id$). A local generator $L_t$ corresponding to this channel is defined via
\begin{eqnarray}
L_t (\rho) = \sum_{i=1}^3 \gamma_i(t) \, (\sigma_i \rho \sigma_i - \rho) ,
\end{eqnarray}
where the relation between probability distribution $p_\alpha(t)$ and local decoherence rates $\gamma_k(t)$ is given by \cite{Chruscinski2013}
\begin{eqnarray}\label{Paulip}
p_0(t) &=& \frac{1}{4}[1 + \lambda_{1}(t) + \lambda_{2}(t)+\lambda_{3}(t)]\phantom{,} \nonumber \\
p_1(t) &=& \frac{1}{4}[1 + \lambda_{1}(t)- \lambda_{2}(t) + \lambda_{3}(t)]\phantom{,} \nonumber \\
p_2(t) &=& \frac{1}{4}[1 - \lambda_{1}(t)+\lambda_{2}(t)-\lambda_{3}(t)]\phantom{,} \nonumber \\
p_3(t) &=& \frac{1}{4}[1 - \lambda_{1}(t)-\lambda_{2}(t)+\lambda_{3}(t)],
\end{eqnarray}
and the time-dependent eigenvalues $\lambda_k(t)$ read
\begin{eqnarray}\label{}
  \lambda_1(t)& =& \exp(-2[\Gamma_2(t) + \Gamma_3(t)]), \nonumber\\
   \lambda_2(t)& =& \exp(-2[\Gamma_1(t) + \Gamma_3(t)]) , \\
    \lambda_3(t)& =& \exp(-2[\Gamma_1(t) + \Gamma_2(t)]), \nonumber
\end{eqnarray}
with $\Gamma_k(t) = \int_0^t \gamma_k(\tau)d\tau$. A Pauli channel is divisible if and only if $\gamma_k(t)$ satisfy the following conditions
\begin{eqnarray}\label{eq:div}
\gamma_1(t) \geq 0, \quad \gamma_2(t) \geq 0 , \quad \gamma_3(t) \geq 0,
\end{eqnarray} for all $t\geq0$.

In a scenario of evolving quantum memory the extractable work is then given by
\begin{eqnarray}\label{pauliex}
W_{ex}(t) &=& [1+ I(S \rangle Q_t)] \, kT \ln 2.
\end{eqnarray}
In this Pauli channel example, the coherent information $I(S\rangle Q_t)$ for the initial  state $|SQ\rangle = \frac{1}{\sqrt{2}}\left(|10\rangle-|01\rangle\right)$ is easily analytically calculated:
\begin{eqnarray}\label{PauliIc}
I(S\rangle Q_t) &=& 1+ \sum_\alpha p_{\alpha}(t) \log_2( p_{\alpha}(t) ).
\end{eqnarray}

In the Article we analyse, and depict in Fig. 2, the time evolution of the extractable work corresponding to two cases of Pauli channels with different values of the $\gamma$-rates. For the first example we choose ${\gamma_1(t)=\frac{\lambda}{2} = \gamma_2(t)}$ and ${\gamma_3(t) =  \frac{\omega}{2} \tan(\omega t)}$ that clearly violate the divisibility condition of Eq. (\ref{eq:div}) and show revivals in the extractable work (see the solid blue line in Fig. 2 of the main article).

For comparison, as the second example we choose the case analysed in \cite{Hall2014, Chruscinski2013}, ${\gamma_1(t)=\frac{\lambda}{2} = \gamma_2(t)}$ and ${\gamma_3(t) = - \frac{\omega}{2} \tanh(\omega t)}$. A Pauli channel with such $\gamma$-rates also leads to legitimate non-divisible evolution whenever $0 \leq \omega \leq \lambda$, however in this case no revivals of the extractable work are present (see the red dashed line in Fig. 2). This is to show that sheer non-divisibility of the evolving channel does not always imply increasing work gain. Interestingly, the second example corresponding to ${\gamma_3(t) = - \frac{\omega}{2} \tanh(\omega t)}$ gives rise to so called P-divisible evolution \cite{PRL-Sabrina} which is characterized by conditions
\begin{eqnarray} \label{P}
\gamma_i(t) + \gamma_j(t)\geq 0,
\end{eqnarray} 
for $i\neq j, i,j =1,2,3$ and all $t\geq0$. As was advocated in \cite{PRL-Sabrina}, a quantum evolution which violates Eq. (\ref{eq:div}), 
but satisfies Eq. (\ref{P}), displays only  weak non-Markovianity. These examples show that essentially non-Markovian evolution, i.e., the one that violates both Eq. (\ref{eq:div}) and Eq. (\ref{P}),  may lead to increase in the extractable work, whereas weakly non-Markovian satisfies $\Delta W_{ex}(t_1,t_2) \leq 0$.

\subsection{Ising Model}

As a second example we consider a qubit (Q or S) transversely coupled to an Ising spin chain in a transverse field \cite{Zanardi}. Such interaction results in a pure dephasing dynamics of the qubit described by the following master equation \cite{PinjaJohn}
\begin{equation}
    L_t (\rho)= i \Lambda(t)[\sigma_z,\rho] + \gamma(t) (\sigma_z \rho \sigma_z - \rho)\ ,
\end{equation}
with $\Lambda(t)$ being the time-dependent Lamb shift and $\gamma(t)$ a time-dependent decay rate further related to the Loschmidt echo $L(t)$ via relation
\begin{equation}
    \gamma(t)=-\frac{\dot{L}(t)}{4 L(t)}.
\end{equation}
Notice that this corresponds to a special case of a Pauli channel with $\gamma_1(t)=\gamma_2(t)=0$ and $\gamma_3(t)=\gamma(t)$. Using the equations (\ref{Paulip}) and (\ref{PauliIc}) it is straightforward to show that
\begin{eqnarray}
I(S\rangle Q_t) &=& 1+\frac{1}{2}\left( 1- \sqrt{L(t)}\right) \log_2\left( \frac{1}{2}\left( 1- \sqrt{L(t)}\right)\right) \nonumber \\
&&+  \frac{1}{2} \left( 1+ \sqrt{L(t)}\right)\log_2\left( \frac{1}{2} \left( 1+ \sqrt{L(t)}\right)\right),
\end{eqnarray} which can be inserted to Eq. (\ref{pauliex}) to obtain the extractable work.
For this model the Loschmidt echo depends on two relevant parameters, $\lambda$ describing the strength of the transverse field and $\delta$ the qubit-environment coupling strength. The analytical expression for the Loschmidt echo is given by
\begin{equation}
L(\lambda, \delta, t) = \Pi_{k>0} \left[1-\sin^2(2 \beta_k) \sin^2(\epsilon_e^k t) \right],
\end{equation}
where $\beta_k$ are the Bogoliubov angles and $\epsilon_e^k$ the single quasiparticle excitation energies of the system with the qubit in the state $|1\rangle$ (for more details see Refs. \cite{Zanardi, PinjaJohn}).

In Fig. 3 of the Article we present three curves describing the time evolution of the extractable work for $\lambda = 1.8$ (orange long-dashed line), $\lambda=0.0$ (green short-dashed line) and $\lambda=0.9$ (black solid line). The remaining parameters are fixed for all the plots as $\delta=0.1$, $J=1$ and the number of spins $N = 4000$.

\subsection{Photonic Band Gap}

The third physical model that we analyse is the amplitude damping channel described by the following exact local generator:
\begin{equation}\label{LAD}
L_t (\rho)=-\frac{i s(t)}{2} [\sigma_+ \sigma_-, \rho]+\gamma(t) \Big(\sigma_- \rho\sigma_+ -\frac{1}{2}\{\sigma_+ \sigma_-,\rho\} \Big),
\end{equation}
with $s(t)=-2 \, \text{Im}\left[\dot{G}(t)/G(t)\right]$ and $\gamma(t)=-2 \, \text{Re}\left[\dot{G}(t)/G(t)\right]$ the time-dependent Lamb shift and decay rate, respectively. In Eq. (\ref{LAD}) $\sigma_+$ and $\sigma_-$ stand for raising and lowering operators, respectively, and the complex function $G(t)$ satisfies $|G(t)|<1$ for all $t>0$, and directly depends on the features of the reservoir. 

The dynamical map of this evolution can be described in terms of the Kraus representation $\Phi_t(\rho) = \sum_{i=1}^2 K_i(t) \rho K_i(t)^{\dagger}$, where 
\begin{eqnarray}
K_1(t) = \left( \begin{array}{cc}  1&0 \\ 0& G(t) \end{array} \right), K_2(t) = \left( \begin{array}{cc} 0&\sqrt{1-|G(t)|^2} \\ 0 & 0 \end{array} \right). \quad
\end{eqnarray}

We consider the model described in Ref. \cite{john}, in which the reservoir is an ideal photonic crystal with a frequency band gap. In such a case the specific form of $G(t)$ is as follows:
\begin{eqnarray}
G(t)&=&2v_1x_1e^{\beta x_1^2 t+i\delta t}+v_2(x_2+y_2)e^{\beta x_2^2 t+i \delta t}\nonumber
\\&-&\sum_{j=1}^{3}a_jy_j[1-\Phi(\sqrt{\beta x^2_j t})]e^{\beta x^2_j t+i\delta t}, \label{gPBG}
\end{eqnarray}
where $\Phi$ is an error function.
One can see that $G(t)$ depends on two relevant parameters, the detuning $\delta=\omega_0 - \omega_e$ from the band gap edge frequency $\omega_e$, which we set to $\delta=-1$, and the characteristic frequency parameter $\beta $ defined as $\beta^{3/2} = \omega_0^{7/2} d^2/ 6 \pi \epsilon_0 \hbar c^3$ with $\epsilon_0$ the Coulomb constant and $d$ the atomic dipole moment.  
In addition: 
\begin{eqnarray}
x_1&= &(A_++A_-)e^{i(\pi/4)},\nonumber\\ 
x_2&= &(A_+e^{-i(\pi/6)}-A_-e^{i(\pi/6)})e^{-i(\pi/4),} \nonumber\\
x_3&= &(A_+e^{i(\pi/6)}-A_-e^{-i(\pi/6)})e^{i(3\pi/4)}, \\
 A_\pm &= &\left[\frac{1}{2}\pm\frac{1}{2}\left[1+\frac{4}{27}\frac{\delta^3}{\beta^3}\right]^{1/2}\right]^{1/3}, \\
 v_j&= &\frac{x_j}{(x_j-x_i)(x_j-x_k)} \quad (j\neq i\neq k;j,i,k=1,2,3), \\
 y_j&= & \sqrt{{x_j} ^2} \quad (j=1,2,3).
\end{eqnarray}

Notice that this dynamical map is not unital, so that one does not have the symmetrical situation with respect to the system and the memory dynamics, that was present in the previous models. Hence, it is a good candidate for investigating the counter-intuitive behaviour of the extractable work described in the Eq. (8) of the Article. 
We therefore focus on the case where the system is subjected to the amplitude damping channel described above. As in previous examples, the initial state is $| SQ \rangle = \frac{1}{\sqrt{2}}( |10\rangle - |01\rangle)$. The entropic quantities entering in the definition of the extractable work can be calculated analytically
\begin{eqnarray}
H(S_t) &=& -\frac{1}{2}\left(2- |G(t)|^2\right) \log_2 \left( \frac{1}{2}\left((2- |G(t)|^2\right) \right)\nonumber \\
&& -\frac{1}{2} |G(t)|^2 \log_2 \left( \frac{1}{2} |G(t)|^2 \right),\\
H(Q) &=& 1,\\
H(S_t Q) &=&- \frac{1}{2}\left( 1- |G(t)|^2 \right) \log_2 \left( \frac{1}{2}\left( 1 - |G(t)|^2\right)\right) \nonumber \\
&&- \frac{1}{2}\left( 1+ |G(t)|^2\right) \log_2 \left( \frac{1}{2}\left( 1 + |G(t)|^2\right)\right),
\end{eqnarray} 
and further used to calculate the mutual information $I(S_t:Q) = H(S_t) + H(Q) -H(S_t Q)$.
Using these we get the analytical solution for the extractable work:
\begin{eqnarray}
W_{ex} = [1-H(S_t) + I(S_t:Q)] \, kT \ln 2.
\end{eqnarray}

This example, where the system is subjected to the photonic band gap amplitude damping, illustrates that the counter-intuitive situation in which one has revivals of extractable work even though the entropy of the system increases, can occur in realistic physical systems; see the shaded intervals of time in Fig. 4 of the main Article.